\documentclass[final,3p,times,authoryear]{elsarticle}

\usepackage{amssymb}
\usepackage{hyperref}
\hypersetup{colorlinks=true,citecolor=blue}
\usepackage{graphicx,amsmath}
\usepackage{ifthen}
\usepackage{multirow}
\usepackage{breakurl}
\def\beq{\begin{equation}}
\def\eeq{\end{equation}}
\def\bea{\begin{eqnarray}}
\def\eea{\end{eqnarray}}

\biboptions{round,comma,sort&compress}

\journal{ArXiv}

\begin{document}
\begin{frontmatter}

\title{The role of money and the financial sector in energy-economy models used for assessing climate policy}

\author[ce]{H. Pollitt \corref{cor1}}
\ead{hp@camecon.com}
\author[RU,CG]{J.-F. Mercure}

\address[ce]{Cambridge Econometrics Ltd, Covent Garden, Cambridge, CB1 2HT, UK}
\address[RU]{Faculty of Science, Radboud University, PO Box 9010, 6500 GL Nijmegen, The Netherlands}
\address[CG]{Cambridge Centre for Environment, Energy and Natural Resource Governance (C-EENRG), University of Cambridge, 19 Silver Street, Cambridge, CB3 1EP, United Kingdom}

\cortext[cor1]{Corresponding author: Hector Pollitt}

\begin{abstract}

This paper outlines a critical gap in the assessment methodology used to estimate the macroeconomic costs and benefits of climate policy. It shows that the vast majority of models used for assessing climate policy use assumptions about the financial system that sit at odds with the observed reality. In particular, the models' assumptions lead to `crowding out' of capital, which cause them to show negative impacts from climate policy in virtually all cases. We compare this approach with that of the E3ME model, which follows non-equilibrium economic theory and adopts a more empirical approach. While the non-equilibrium model also has limitations, its treatment of the financial system is more consistent with reality and it shows that green investment need not crowd out investment in other parts of the economy -- and may therefore offer an economic stimulus. 

The implication of this finding is that standard CGE models consistently over-estimate the costs of climate policy in terms of GDP and welfare, potentially by a substantial amount. These findings overly restrict the range of possible emission pathways accessible using climate policy from the viewpoint of the decision-maker, and may also lead to misleading information used for policy making. Improvements in both modelling approaches should be sought with some urgency -- both to provide a better assessment of potential climate policy and to improve understanding of the dynamics of the global financial system more generally.

\end{abstract} 




\end{frontmatter}


\section{Introduction}

\subsection{We can meet the 2$^{\circ}$C target-- but who will pay?}

There is a gradually emerging consensus that a global emissions pathway that is consistent with the target of keeping emissions concentrations below 450~ppm, and thus of having a 50\% chance of limiting anthropogenic climate change to 2$^{\circ}$C above pre-industrial levels, is technologically feasible \citep{IPCCAR5WGIII}. The question of whether the 2$^{\circ}$C target will be met or not is therefore a political one to do with the allocation of scarce resources -- essentially to determine who will pay if the world is to meet its collective target.

It seems beyond doubt that targets for emissions levels will not be met without the introduction of new policy. As outlined in \cite{Grubb2014}, there are three main forms this policy could take:

\begin{enumerate}
\item Policies to improve the use of energy with existing technologies, such as enforcing efficiency standards through regulation.
\item Policies to ensure an efficient allocation of resources given existing technologies, for the main part through market-based mechanisms (demand-pull policies).
\item Incentives to develop new technologies, for example through providing tax credits on R\&D expenditure (supply-push policies).
\end{enumerate}

These policies differ substantially in scope, and their responsibility may not even fall under the same government departments, but they do have some common characteristics. All will involve a reallocation of economic resources compared to what would have happened without government intervention. Most will involve substantial amounts of investment. This means that the policies will have impacts on both the real economy and across the financial system; understanding the interaction of investors in low-carbon technologies with the banks that provide the necessary credit and the companies that produce or install the equipment will be key to assessing overall impacts.

In summary, all of these types of policy will lead to economic winners and losers, with financial consequences at both the micro and macro levels. In a modern economy, all must therefore be justified prior to implementation. Quantitative models contribute to this process by providing evidence of the likely costs and benefits of potential policy.

\subsection{The role of E3 models and IAMs in policy analysis}

The emphasis placed on computer modelling in climate policy has been increasing steadily as data have improved and additional computer power has allowed the development of more complex tools. Large-scale climate models and Integrated Assessment Models (IAMs) are central to the analysis carried out by the IPCC both to estimate the current emissions trajectory and paths with which there is a reasonable chance of staying within the 2$^{\circ}$C target. 

When it comes to assessing the implications of climate policy on the wider society, E3\footnote{By this we mean models where all the main macroeconomic national accounting variables are endogenous outputs from the model so, for example, pure energy systems models are generally excluded.}  (Energy-Environment-Economy) models are applied to estimate impacts on indicators such as GDP, welfare and employment. As is often the case, the terminology is not always used consistently but here we define E3 models as essentially macroeconomic models that have been extended to include some physical relationships. Their use has been well-established since at least the IPCC's second assessment report \citep{IPCC1995} and the relative weight placed on model results has increased over the past decade. For example, the European Commission's Impact Assessment guidelines states that for any policy assessment or evaluation:

\begin{quote}
You should keep in mind that the credibility of an IA depends to a large extent on providing results that are based on reliable data and robust analysis, and which are transparent and understandable to non-specialists. This exercise will usually require an inference from the collected data, either formally through statistical analysis or model runs, or more informally by drawing on an appropriate analogy with measured impact or activities. This assessment should go beyond the immediate and desired aspects (the direct effects) and take account of indirect effects such as side-effects, knock-on effects in other segments of the economy and crowding out or other offsetting effects in the relevant sector(s) \citep[][ p. 32]{EC2009}.
\end{quote}

And also that:
\begin{quote}
If quantification/monetisation is not feasible, explain why \citep[][ p. 39]{EC2009}.
\end{quote}

Taken together, and given the often disparate effects of climate policy on the economy, the message is quite clear -- for any new climate policy proposals to be accepted at European level it is necessary to provide model-based evidence of the macroeconomic impacts.

\subsection{Different types of macroeconomic models}

In many cases policy makers' understanding of macroeconomic models has not kept pace with the more prominent role that the models play in policy analysis. This is unfortunate as it is not possible to properly interpret the results from the models without understanding the underlying mechanisms; and, furthermore, there are substantial differences between the ways the models work. It is recognised in the field that there is an inherent difficulty in communicating an understanding of complex tools to time-pressured policy makers who may not come from a quantitative or economic background. There are efforts to address this, for example in providing specialised training.

The models that are used to assess the macroeconomic impacts of climate policy fall broadly into two groups. These are:\footnote{There are also some models that fall between these two definitions, although their treatment of finance will generally follow the neoclassical approach. Small-scale Integrated Assessment Models (IAMs) such as DICE also fall into this category.}
\begin{itemize}
\item Computable General Equilibrium (CGE) models that are usually described as being based on neoclassical microeconomic assumptions. These models assume that agents (e.g. firms, households) optimise their behaviour so as to maximise their personal gains. Well-known international CGE models include GEM-E3 \citep{Capros2013}, GTAP \citep{Hertel1997} and the Monash model \citep{Dixon2002}. The Handbook of Computable General Equilibrium Modeling \citep{Dixon2013} describes in detail how these models work. Model intercomparison exercises, such as those carried out by the Energy Modelling Forum \citep[e.g.][]{Weyant2006} typically compare the results from different CGE models.

\item Macro-econometric models that are derived from a post-Keynesian economic background.\footnote{Again there is a potential issue with terminology here. By macro-econometric models we mean those that are derived from a post-Keynesian economic background. This category does not include CGE models that have parameters which have been estimated using econometric techniques.} These models do not assume that agents optimise their behaviour, but instead derive behavioural parameters from historical relationships using econometric equations (which allow for `bounded rationality'). Well-known international macro-econometric models include E3ME \citep{E3ME} and GINFORS \citep{Lutz2009}.

\end{itemize}

The aim of this paper is not to describe in detail the differences between the modelling approaches.\footnote{Pollitt et al (2015) and Knopf et al (2013) expand on this in the context of climate policy; Cambridge Econometrics et al (2013) provides a practical example in relation to EU policy to meet the 2∞C target.} The focus of this paper is instead on describing how the different models represent the global investment that will be required to meet the 2$^{\circ}$C target, how such representations influence model results, and how this information can be interpreted by decision makers. Closely tied to this issue is the question of how the models treat banks, money and the financial sector, which we introduce below. 	

\subsection{Why is the treatment of money and finance important in macroeconomic models?}

It is beyond doubt that substantial investment will be required to meet the 2$^{\circ}$C target. The \cite{IEAWEO2014}, p93, estimates that at global level \$2.4trn (2013 prices) `clean energy investment' must be made annually in its 450~ppm scenario. All of this investment must be financed somehow; although some could be diverted from investment in developing fossil fuel resources, the investment-intensive nature of low-carbon technologies (e.g. renewables, energy efficiency) means that any policy scenario in which emissions are reduced is likely to require an increase in energy-sector investment. The question of how the investment is financed, and whether more investment resources can be mobilised, is therefore key to understanding the economics of a low-carbon transition.

There are, however, also other reasons to focus attention on finance. As was made painfully aware by the financial crisis and subsequent recession, even sophisticated macroeconomic models have only a rudimentary treatment of finance.\footnote{See \cite{Keen2011} for a detailed general discussion. See \cite{Anger2015} for a recent shorter discussion in the context of climate policy.} While there have been attempts outside mainstream economics to build macroeconomic models with better links to finance \citep{Minsky1982}, these are not developed enough to apply to climate policy.\footnote{One potential tool that has been developed is the Global Policy Model \citep[see][]{UNDESA2009}, although it does not go into sectoral detail.} The treatment of banks and the financial sector is therefore done largely by assumption. Furthermore, as we shall demonstrate, these assumptions vary enormously between the different modelling approaches.

These modelling approaches are described in Section 3. First, however, we describe the underlying theory and how it relates to the different schools of economic thought. In Section 4 we turn attention to the lessons for policy makers from our analysis. Section 5 concludes.

\section{Money and the financial sector in the different schools of economic thought}

\subsection{Introduction}

The focus of this paper is on the impacts of assumptions in macroeconomic modelling approaches, and thus it is necessary to have a basic understanding of the underlying theory and philosophy in order to understand how the models work. This section therefore gives a brief overview of how finance is treated in the most relevant schools of economic thought.

\subsection{Money and finance in neoclassical economics}

Most readers will be familiar with the Efficient Markets Hypothesis (EMH) that forms the core of financial theory in neoclassical economics. The EMH postulates that markets are `efficient' and that the prices that are set accurately reflect all of the available information. The underlying assumptions, such as the same information being available to all individuals who act rationally, are broadly, if not entirely, consistent with those used in CGE models. Although the EMH has been criticised heavily for making these assumptions, especially since the global financial crisis, it is still the standard approach that is taught in economics textbooks.

While the EMH is certainly relevant to the modelling of energy and climate policy, for example in the way that optimal carbon prices are determined in most modelling approaches, neoclassical theory of the money supply is much more important in determining the macro-level impacts of climate policy in CGE models (e.g. GDP, welfare) -- as we shall show in the next section. In neoclassical theory the money supply is effectively determined by the central bank which, in modelling terms, makes it exogenous. If there is an increased demand for money from commercial banks, government or private sector institutions, interest rates (i.e. the price of money) will adjust in response and there will be no change in the overall supply of money. 
Furthermore, if the central bank does increase the money supply (in nominal terms) this does not have any impacts on real rates of economic activity. Instead, prices and inflation rates automatically adjust by the same relative amount, an approach that is consistent with the optimisation principles applied in the modelling; if all available resources are being used optimally already then making more money available will just lead to higher prices for these resources. Within economics this theory is referred to as the neutrality of money.

\subsection{Money and finance in New Keynesian economics}

New Keynesian economists (not to be confused with post-Keynesian economists) also accept money neutrality in the long run, but allow for changes in the money supply to have real impacts in the short run. This is because under New Keynesian assumptions interest rates are set by the central bank rather than the market and there are time lags in adjustments in prices; these are reflected in the Dynamic Stochastic General Equilibrium (DSGE) models that are used by central banks and can include an explicit representation of money.

Nevertheless, the long-run outcomes will be the same as those described above, with prices adjusting to equilibrium value. As climate policy analysis typically focuses on long-run outcomes, and DSGE models are not commonly used assessing climate policy, we do not develop this further in the present paper.

\subsection{Money and finance in post-Keynesian economics}

Money plays a central role in post-Keynesian economics -- as noted in \cite[][p18]{King2015}, the term appears in the full title of Keynes' \emph{General Theory}; in the recent textbook by \cite{Lavoie2014}  money and finance are introduced before the real economy. In contrast to the neoclassical approach, post-Keynesian economists follow a theory of `endogenous money'.\footnote{There is some contention regarding whether Keynes himself supported the theory of endogenous money, see \cite{Dow1997} for a discussion.} The approach is based on the fact that in a modern economy, most of the money is created by commercial banks through the advancement of new loans. Due to leverage effects,\footnote{The act of using deposits to finance multiple loans simultaneously.} banks do not need to receive additional deposits to make new loans. 

When the banks do make new loans, they create simultaneously a matching deposit in the borrower's bank account, thereby creating new money. \cite{McLeay2014} provides a very clear summary of the processes involved but, in summary, the volume of loans and therefore money supply is at least in part determined by broader macroeconomic conditions (i.e. whether the banks see profitable commercial opportunities). Depending on whether reserve requirements exist, and their magnitude, central banks are assumed to print `on demand' the amount of money required by the commercial banks in order to underwrite these loans.\footnote{In practice, central banks increase the size of deposits (and liabilities) of commercial banks held at the central bank.} In advanced economies this is what the central banks generally do.

In post-Keynesian economics the money supply is also important because it leads to real economic effects, particularly in the short run but potentially with long-run impacts. As prices do not adjust instantly (or even at all), providing more money to make purchases can lead to an increase in aggregate demand, pulling previously unused resources into the system. It is in this way that interest rate policy is applied (encouraging banks to make loans that would stimulate aggregate demand). Quantitative easing follows a similar idea, although in both cases the banks must be willing to make the loans if demand is stimulated. While the New Keynesian theory described above also favours these policies, post-Keynesian economists stress features such as uncertainty and path dependence, i.e. that short-term developments can influence long-term outcomes.

The post-Keynesian economists are themselves divided into two groups: horizontalists and structuralists \citep{Pollin1991}. The names reflect the shape of the money supply curve (in neoclassical theory it is vertical, i.e. fixed). If the money supply curve is horizontal, banks are free to lend infinitely without restriction; the sole limit on lending is thus the degree of profitable opportunity \citep{Moore1988}. Interest rates are exogenous under this approach, although some variants with upward sloping curves have been suggested. Structuralists endogenise the interest rate by adding further real-world factors, but at the expense of a more complex theory \citep{Palley2013}. 

Of crucial importance is that the empirical evidence supports the post-Keynesian formulation. \cite{Anger2015}, p183 lists numerous studies that show the limited role of the central bank in controlling the amount of credit available, and therefore the supply of money to the economy. \cite{Arestis2011} conclude that ``[...] the analysis of macroeconomies cannot be reduced to studies of economies without money and finance'' -- as noted in the introduction above, this finding is particularly relevant to policies that are designed to promote investment, including climate policy.

\subsection{Money and finance in the post-Schumpeterian (evolutionary) school}

Money creation is also a key component of the post-Schumpeterian school, also known as evolutionary economics. In \cite{Schumpeter1934,Schumpeter1939}, productivity growth is effectuated by the entrepreneur, who has no funds but has ideas. He innovates in the production process by inventing new combinations of resource use (\emph{ibid}) which increase productivity and/or lower costs. If he is successful, innovation confers him a temporary monopolistic profit, until the time when the competition catches up, at which point economy-wide prices have declined. In the long run, this generates economic development. To carry this out, the entrepreneur needs finance, however, which is provided for by banks, based on the credibility of his business plans, past successes, and a general confidence in the economy. Banks create loans, which the entrepreneur pays back with his profits. In a volume published posthumously, \cite{Schumpeter2014} gives a complete positive theory of the money supply and of the role of banks, which is, in its broad lines, equivalent to that of the post-Keynesian school.\footnote{Schumpeter wrote his \emph{Treatise on Money} during his career, ultimately deciding not to publish it, which has left his view on the monetary system obscured. This work has recently been published in German in 2008, and translated to English in 2014. The theory involved in his \emph{Theory of Economic Development} and \emph{Business Cycles}, however, already implies directly the process of money creation, in fact more clearly so than Keynes' \emph{General theory} does. This can also be appreciated in later work of the evolutionary school.}

In this perspective, Schumpeter's view of the economic process reveals another side of the same coin, in comparison to the post-Keynesian view: money and finance come first, and productivity growth and economic activity follow. That perspective implies, equally to Keynes', that the economy is demand-led. However it also adds a new micro perspective: that of the entrepreneur. Effectively, money creation comes with trust, by finance institutions, that the entrepreneur's innovation will generate profits and that loans will be paid back with interest. Thus expectations are a key feature of the model.

A second key element is added by Schumpeter in his work on business cycles: technological change comes in waves, and innovation clusters. It can readily be seen that if, for any reason, innovative activities are linked to one another in some way (e.g. constellations of innovations related to the steam engine), and limited in extent of development (e.g. covering Britain with railways), then limited waves of activity arise in different sectors at different times, corroborated to waves and slumps in investment. Thus the \emph{clustering of innovative activity} leads to \emph{economic cycles} of different lengths. Additionally, the success of some innovations may lead to euphoria in the financial sector, while the depletion of innovative possibilities leads to pessimism, and thus bubbles can arise (for example, the dotcom bubble).

Throughout his work, Schumpeter insists on adopting a historical approach. Following this, work has been done on characterising the great historical waves of innovation \citep{Freeman2001,Freeman1988}, and matching them to the structure of investment fluctuations and technology-related financial crises over history \citep{Perez2001}. Meanwhile, the network nature of the clustering of related innovations has been studied computationally by \cite{Arthur2006}. These analyses provide a structure to the process of clustering of investment related to productivity growth, and thus to business cycles. While such insights cannot readily be brought into economic models without prior descriptions of the technologies in question, they suggest a very clear methodology for how to study given problems of technological change, including notably the challenges involved in climate change mitigation.

\section{Money and the financial sector in the different modelling approaches}

\subsection{Introduction}

The previous section showed that the treatments of money and finance vary considerably between the different strands of economic theory. In this section we turn attention to the practical application of these theories through computer models. The explanations focus on the assessment of energy and climate policy but it would be possible to draw the same conclusions for any type of policy that was investment intensive and for which we were interested in the long-run outcomes.

Our review is carried out at the global level. International financial flows can complicate the issue when considering policies at regional level, due to interactions with exchange rates and international trade. The aim of this paper, however, is to provide a basic understanding of the most important differences between the modelling approaches, so we focus on the simpler case.

It is important to note that all the different models we look at observe the macroeconomic identity that savings should equal investment but, as we shall see, they have quite different interpretations of how the balance is met. First we describe the processes involved in the neoclassical CGE models before explaining the roles of money and finance in the post-Keynesian E3ME macro-econometric model. The table towards the end of this section summarises the key differences between modelling approaches and the section finishes with an example relating to climate policy. 

\section{The role of money and finance in CGE models}

The sixth part of Walras' Elements of Pure Economics \citep{Walras1874,Walras1954}, widely regarded as the bible for CGE modelling,\footnote{e.g. ``Walras not Keynes is the patron saint of CGE models'', Robinson and Devarajan (2012), p282.} is titled `Theory of Circulation and Money'. This title hints at the role of money in the economy in CGE models -- as a means to allow the transactions of goods and services. Lessons 28 and 29 of the book expand on the approach; the demand and supply of money are described in micro terms, with money being held to allow the immediate purchases of goods and services -- e.g. for consumers:

\begin{quote}
[...] a  certain quantity of cash on hand and savings which are mathematically determined by the same attainment of maximum satisfaction, under the same aforementioned conditions, in accordance with each consumer's initial quantity of money and not only his utility or want functions for the services of availability of new capital goods \underline{in the form of money} rather than in \underline{kind;} \citep[][p316, emphasis in original]{Walras1954}
\end{quote}

A similar definition for producers is provided on page 317. The passage goes on to describe an agent's cash balance as:
\begin{quote}
[...] not only in order to replenish these stocks [final products] and make current purchases of consumers' goods and services for daily consumption while waiting to receive rents, wages and interest payable at fixed future dates, but also in order to acquire new capital goods. \citep[][p317]{Walras1954}
\end{quote}

After explaining the reasons for holding cash, Walras describes the role of lending and borrowing money in the economic system:
\begin{quote}
That is not all [...] \underline{Capital} being defined as ``the sum total of fixed and circulating capital goods hired, not in kind, but in \underline{money}, by means of \underline{credit}'' [...] This quantity of repaid capital, to which land-owners, workers and capitalists add a certain excess of consumption over income, or from which they subtract a certain excess of consumption over income, constitutes the day-to-day amount of savings available for lending in the form of money. \citep[][page 317, emphasis in original]{Walras1954}
\end{quote}

The implications are quite clear -- savings are defined as the difference between consumption and income and this difference is equal to the sum of money available for lending, i.e. a change in lending must be compensated by a change in current consumption. The price of money is adjusted so as to obtain equilibrium in the market for money (p327), there is no explicit role for the banking sector and it is assumed that either savers and borrowers interact directly or the banks act as frictionless paths in which money is channelled between the savers and borrowers. Furthermore, risk on investment is not part of the theory, and all of the money available for lending is always fully used.

Walras also discusses the impact of changes in the money supply (p327-329). The assumption is that the value of money is only determined by the value of the goods and services it may purchase and hence is ``inversely proportional to its quantity'' \citep[][p329]{Walras1954}. Interestingly, even in the 19th century Walras was aware of the restrictive assumptions relating to price adjustments that were necessary to justify this proposition (p328); he describes the treatment as one of `almost rigorous exactness'.

The treatment of money in modern CGE models is based on the approach described by Walras, with the total money supply fixed in real terms and money used as a means of exchange rather than something that can have an impact on rates of real economic activity. The current handbook \citep{Dixon2013} pays little attention to money. A search for the word `money' reveals first a description of the MAMS model that includes:

\begin{quote}
Like most other CGE models, MAMS is a `real' model in which inflation does not matter (only relative prices matter). Given this, there is no significant gain from having a separate monetary sector. \citep[][p234]{Lofgren2013}
\end{quote}
Then a similar paragraph for the 1-2-3 model:
\begin{quote}
It [the exchange rate] can be seen as a signal in commodity markets and is in no sense a financial variable since the CGE model does not contain money, financial instruments or asset markets. \citep[][p281]{Robinson2012}
\end{quote}
Only the G-Cubed model description \citep{McKibbin2012} talks about money at length, although it must be noted that it is in the context of non-equilibrium (i.e. it is not pure CGE). The other search results (excluding DSGE descriptions) refer to money as a metric for presenting model results rather than something that can impact on these results.

\section{The role of money and finance in the post-Keynesian E3ME model}

E3ME is a global macro-econometric model. It combines input-output analysis with sets of econometric equations that determine the components of aggregate demand and price levels. The basic economic framework is extended to incorporate physical flows of energy use, materials consumption and greenhouse gas emissions. The model's parameters are derived from time-series historical data and it projects forwards annually to 2050.

The role of money and the financial system is not covered directly in the current version of the E3ME model manual \citep{E3ME}, although there are several references to related features. Essentially the model provides a `horizontalist' approach (see previous section) to banks, lending and the money supply. Investment is determined by expectations of future output levels, which are based on current activity. The interest rate is fixed as exogenous and there are no set limits on the amounts that banks can lend. It should be noted that this does not mean that banks are allowed to lend completely freely as past regulations will be factored into the model's econometric parameters which are derived from historical data. However, overall the implication is that an improvement in economic conditions will lead to an increase in the demand for money which will be followed by an increase in the money supply. This means that an increase in investment does not need to be financed by an increase in savings or reduction in investment elsewhere.\footnote{  It is, of course, possible to specify a source of financing instead of assuming that the investment is financed by borrowing. When assessing policies directly additional assumptions are usually added \citep[see e.g.][]{CamEcon2013}. For example, public expenditure on energy efficient equipment is usually assumed to be financed by higher tax rates (i.e. enforcing savings on workers). Investment in renewables by the power sector may be financed by charging higher electricity prices to consumers.} Instead, an increase in investment can be financed by an increase in public and/or private debt -- and the savings-investment identity is maintained through an expansion of the economy that generates additional savings. Or, to put it another way, capital markets do not enforce a crowding out of investment.

This feature would possibly not matter if the model embodied other equilibrium properties. For example, if full employment were obtained, there would be no additional workers left who could be employed to build any new equipment or infrastructure. However, the demand-driven nature of the model means that this is not usually the case, and unemployment exists in E3ME. So an increase in investment levels can lead to an overall increase in economic activity rates. This sets the model apart from the results that are typically obtained by CGE models.

The treatment raises the question of what the capacity constraints in the model are. Would it be possible to keep on financing higher levels of investment through ever increasing levels of debt? The answer to this question is somewhat mixed. First, as the public sector is treated as exogenous in the model it is up to the model user to enter assumptions, including on borrowing rates; if these assumptions are not realistic then the model responses will not be either -- although of course there is considerable uncertainty about how much a country can borrow under different economic conditions (as has been highly evident since the 2008 global financial crisis). Second, there is one obvious capacity limit imposed by the stock of available labour. While involuntary unemployment is a standard endogenous feature of the model, if the economy moves towards full employment then wages increase and labour market crowding out occurs.

More generally, the modeller faces the same challenges as economists who try to measure the `output gap'\footnote{This issue can be overstated. In the wider literature the unemployment rate is often used as a proxy for the output gap and unemployment is one of the model outputs. However, for climate policy scenarios it is possible that there will be other specific capacity constraints (e.g. a material input or skill group) that limit production levels.}  or the potential levels of debt that individuals, firms or national governments are willing or able to take on. There is considerable uncertainty over the level of any economy's actual capacity to produce a complex range of goods and services to meet global demands, at present and even more so in the future. For this reason, expected future capacity is modelled implicitly through econometric equations that take into account past growth rates. If a sector's output increases more quickly than expected, it will increase prices and there may be import substitution. Workers in that sector may also work longer and receive higher wages (e.g. through bonus payments). In summary, inflationary pressures will start to build at a time of rapid economic expansion.

\begin{table}[t]
\begin{center}
	\begin{tabular*}{1\columnwidth}{ l|l|l }	
							&Standard CGE model & E3ME model 		\\
		\hline
		\hline
		Theoretical background	&Neoclassical			&Post-Keynesian	\\
		Money Supply			&Exogenous			&Endogenous		\\
		Capital market crowding out	&Yes -- full		&No				\\
		Price adjustment		&Instant				&Dynamic over time	\\
		Money neutrality		&Yes					&No				\\
		Capacity constraints		&Current production		&Implicitly given, based on recent history \\
		Labour crowding out		&Yes -- full			&Partial, increasing towards full employment \\
		Other crowding effects	&Yes -- full			&Only in short term, if growth exceeds trend rates \\
		\hline
		\end{tabular*}
	\caption{Key characteristics of the models discussed in the text.}
	\label{tab:table}
\end{center}
\end{table}

\subsection{Summary of features of the models -- and a worked example}

Table \ref{tab:table} summarises the key findings of the sections above. To illustrate these features, we can take the example of a carbon tax with revenue recycling (e.g. through reduced income/corporate taxes). If the policy is revenue neutral overall, then in a CGE model we are considering a reallocation of resources -- a new optimal point in sectoral prices space -- which will take us away from the optimal economic starting point and therefore reduce rates of economic activity. But in the E3ME framework there is the possibility that the policy stimulates additional investment financed by borrowing, in which increasing debt levels contribute to aggregate demand \citep[as described in Chapter 12 of][]{Keen2011}, drawing upon unused economic resources to increase overall production levels. Higher current rates of output will lead to expectations of higher future rates of output, and there is a long-run increase in production levels which will be used to pay down the initial borrowing. This results in higher GDP and (possibly) employment levels \citep[see e.g.][]{Barker2015}. 

In fact, in E3ME there need not be an increase in low-carbon investment following such a carbon tax. If consumption of oil and gas falls then the income of oil and gas exporting countries will also fall. As the saving ratios of oil exporters are typically higher than the saving ratios of fuel consumers (e.g. compare sovereign wealth funds to households with cars), then there can be a similar stimulus-type effect: net global debt is still higher but through reduced saving rather than increased borrowing. We can see this in reality when there are reductions in the global oil price; even if the consumption of oil does not change there can be a noticeable boost to global GDP growth rates due to a shift in global saving rates,\footnote{Although it should also be noted that there can be marginal effects causing reductions in global demand -- i.e. firms and governments that lose income may be quicker to cut back spending than consumers will be to spend their extra income.} although the GDP of the major oil and gas exporting countries will fall. In a CGE model, the basic reallocation of resources without changes in net saving would not produce this result.

\subsection{A post-Schumpeterian view: the entrepreneur borrowing to invest in new technology}

Modelling climate change mitigation is often done from a `bottom-up' technology perspective, and this provides an opportunity to study its relationship with the economic process. Indeed, while in E3ME the process of money creation for productivity growth is implied but not described explicitly, connecting explicit models of technological change to E3ME, with investment and price feedbacks, enables us to do precisely that: studying the process of money creation for financing technology ventures. It furthermore adds a clear post-Schumpeterian perspective to modelling through an explicit representation of the entrepreneur seeking finance to invest in technology. In a context of climate change mitigation, this is a crucial aspect to explore \citep{Lee2015, Mercure2015b}.

Indeed, for purposes of studying emissions reduction policies, an evolutionary model of technology selection and diffusion was recently connected dynamically to E3ME, the only one of its kind, named `Future Technology Transformations' \citep[FTT, ][]{Mercure2012, Mercure2014}. In its application to the electricity and transport sectors, this model explores the impact of policies on the choice of investors or consumers for technologies that provide various societal services (electricity and mobility in this case), of which the demand is evaluated econometrically by E3ME. As noted above, low-carbon technologies tend to be, in general, more capital intensive than incumbent fossil fuel systems \citep[see also][]{IEAWEO2014,IEAWEIO2014}. In scenarios where policies incentivise investors to adopt new technologies, the additional investment, in comparison to a baseline, does not `crowd-out' investment elsewhere: money is created to finance these ventures. However, given the assumption that banks consider these ventures profitable, the self-consistent structure is adopted where firms pass on higher financing costs to consumers, for example with a higher price of electricity.

This therefore creates a modelling system that contrasts quite starkly with equilibrium approaches: money is lent by banks for financing evolving technology developments, which are paid back through higher receipts from consumers over the lifetime of the new capital. Therefore, in contrast to equilibrium approaches where \emph{economic costs are incurred first, and benefits may arise later, here, economic benefits arise first, and costs are incurred later.} For instance, if high investments are made early to radically transform the electricity sector to reduce emissions (e.g. replacing coal power in China), high increases in employment and income may temporarily take place. However, when the technological transformation is completed, significant levels of debt may remain, which leaves society to live with a legacy of debt servicing payments, of which the costs are given to consumers through prices. Depending on whether this transformation has enhanced productivity growth and/or international competitiveness, the transformation may be beneficial or detrimental to economic performance in the long run.\footnote{If learning-by-doing cost reductions and/or cross-sectoral learning spillovers produce a permanently lower operation cost for the electricity system and/or lower cost of living, Schumpeter's description of economic development is then fully realised.} 

\subsection{Implications for policy makers}

So far in this paper we have outlined the following:
\begin{itemize}
\item Macroeconomic models are used frequently to estimate the economic costs and benefits across a range of policy areas, including climate policy.
\item Much climate policy (e.g. renewables, energy efficiency) requires substantial investment and financing for this investment.
\item The treatment of finance varies considerably between modelling approaches and (especially) in the most common CGE approach it is largely constrained by assumption.
\end{itemize}

There is therefore a difficult task for policy makers to interpret and compare model results, given the differences in treatment of money, the financial system and investment, much of which is relatively undocumented. The key result in economic terms, however, is that:

\emph{In a CGE model, an increase in investment due to climate policy will always mean either a reduction in investment elsewhere in the economy or an increase in savings at the expense of current consumption, due to financial crowding out effects. Due to diminishing marginal returns, this reallocation of investment is effectively certain to have a negative effect on total economic production levels.}\footnote{  Assuming, as is usually the case, that climate impacts are not included in the modelling exercise -- if the avoided costs of climate change were included, it could be possible to still get a positive result.}

\emph{In contrast, in a non-equilibrium macro-econometric model, if investment opportunities are sufficiently commercially attractive, banks may choose to increase their lending, leading to an increase in net credit and the broad money supply, in turn stimulating real economic activity and leading to higher rates of output and employment. While in the longer term there may be costs as loans are repaid, higher rates of production can stimulate further activity, meaning long-term impacts need not be negative.}

To put it another way, a CGE modelling approach represents a worst-case outcome for policy makers; the starting point is one of optimal use of resources (including in the financial sector) from an economic perspective and the policy shows the negative impacts of intervention and a reallocation of limited resources. This raises the question of how close we are to an optimal starting point? In 2015 the answer seems to be `not very' with a combination of economic recession, demographic change and a shift to less capital-intensive industries leading to a persistent `global savings glut' \citep{Zenghelis2011}. The continuation of Quantitative Easing (QE) in Europe, a policy designed to increase the money supply directly without the intermediation of banks, suggests a position that is far from optimised. However, while all of this suggests benefits from encouraging investment in the short term, climate policy scenarios usually consider the period out to 2030 and beyond; a wide range of possible outcomes can be predicted for macroeconomic conditions so far ahead.

The simulation-based approach offered by the non-equilibrium macro-econometric model is much more in line with how the financial system works in most countries but is by no means perfect. It does not present a best-case outcome but, by not having an explicit treatment of possible financial constraints, it is more likely to be erring on the optimistic side. Indeed, the power of decision for the creation of loans belongs to banks, and banks can at any point refuse to lend. It would, therefore, be desirable to test the sensitivity of model results to the addition of constraints (e.g. by adjusting interest rates, or changing baseline unemployment rates) to see how important the assumptions are -- a similar exercise could be carried out with a CGE model but the model would become difficult to solve in non-equilibrium conditions.
One possible solution to the problem is to use both modelling approaches to test climate policy, as is now common within the European Commission \citep[e.g.][]{CamEcon2013}. Although this clearly requires additional resources for policy analysis, there are benefits both in terms of obtaining a range of results but also in the discussion between the model results, which can help policy makers to understand some of the key assumptions that are involved (including the treatment of money and finance).

\section{Conclusions}

If the world is to meet the 2$^{\circ}$C target, it is clear that substantial levels of additional investment will be required. How this investment is financed is a key question for policy makers, as has become clear from the UNFCCC negotiations in recent years.

In attempting to assess the costs and benefits of climate policy (and also other policy areas), policy makers now frequently turn to macroeconomic models to provide estimates. However, as shown in this paper, the majority of these models make the assumption that the investment can only be financed by taking investment from elsewhere in the economy, or by reducing current consumption (and welfare) levels. This is not consistent with how the financial system works in the real world, as demonstrated by the real-world use of interest rate policy and QE, which would have no impact in these models.

The alternative approach, offered by the relatively few models that follow post-Keynesian principles, is not without limitations either but offers a version of the financial system that is closer to that which we can observe. However, questions of economic capacity which have only a limited representation in the model place a burden on the model operator to ensure that policy scenarios are realistic.

In summary, the analysis in this paper shows that the post-Keynesian approach appears to be the best that the research community can offer at present but both modelling approaches could and should be improved further. On the surface, it looks like the post-Keynesian modelling approach is in a better position to adapt, as assumptions about optimisation and fixed supply are fundamental to the CGE approach, which lends it significantly less flexibility.

Thus, when assessing policy, decision makers and policy analysts must identify what the implications are of their choice of models and associated underlying assumptions. The simultaneous use of models with different theoretical underpinnings allows for safer identification of possible ranges of economic outcomes. 

Given how important issues of finance are in estimating the impacts of climate policy in the models, our view is that improving the treatment of finance in the models should be given priority in coming years -- the benefits would be a more accurate representation of the impacts of climate policy and investment across the world's economies more generally.

\section*{Acknowledgements}

The authors thank T. Barker and R. Lewney for highly informative discussions. J.-F. M. acknowledges funding from the UK EPSRC fellowship no EP/ K007254/1.

\section*{References}

\bibliographystyle{elsarticle-harv}
\bibliography{../../CamRefs}

\end{document}